\documentclass[aps,prl,twocolumn,showpacs]{revtex4}
\usepackage{amsmath}
\usepackage{graphicx}

\usepackage{epsf}

\newcommand \bea {\begin{eqnarray} }
\newcommand \eea {\end{eqnarray}}

\newcommand{\beg}{\begin{equation}}
\newcommand{\en}{\end{equation}}

\newcommand{\eps}{\varepsilon}
\newcommand{\lam}{\lambda}

\newcommand{\re}[1]{(\ref{#1})}

\begin{document}

\title{Spectroscopic signatures of nonequilibrium pairing in atomic Fermi gases}

\author{M. Dzero$^1$, E. A. Yuzbashyan$^1$,
B. L. Altshuler$^2$ and P. Coleman$^{1}$}
\affiliation{$^1$Center for Materials Theory, Rutgers University, Piscataway, NJ 08854, USA\\
$^2$ Department of Physics, Columbia University, New York, NY 10027, USA}

\begin{abstract}
We determine the radio-frequency (RF) spectra for non-stationary   states of a
fermionic condensate produced by a rapid switch of the
scattering length.  The RF spectrum of the nonequilibrium state with constant  BCS order parameter has two features in contrast to equilibrium where there is a single peak.  The  additional feature reflects the presence of excited pairs  in the  steady state.  In the state characterized by periodically oscillating order parameter RF-absorption spectrum
contains two sequences of peaks spaced  by the frequency of oscillations. Satellite peaks appear due to a process where an RF photon in addition to breaking a pair emits/absorbs oscillation quanta.

\end{abstract}


\maketitle

Cooper pairing in ultra-cold  Fermi gases has been a major focus of research in the past few years. Remarkable experimental techniques such as sweeps across the Feshbach resonance \cite{Regal2004,Zwerlein2004}, generation of collective modes
\cite{Bartenstein2004} and vortex lattices\cite{Zwerlein2005}, and radio-frequency (RF) spectroscopy \cite{Chin2004,Schunck2007}
have been developed to probe the paired state. While it was crucial to establish for cold
gases well-known signatures of fermionic pairing, of a key interest are regimes not easily accessible in superconductors, e.g.  strong interactions in the vicinity of the Feshbach resonance and highly imbalanced mixtures.

One of the most interesting possibilities is to access the non-adiabatic coherent dynamics of fermionic condensates [\citealp{Galaiko1972}--\citealp{Emil2006}]. Driven out of equilibrium by a sudden change of the pairing strength on the BCS side of the Feshbach resonance, these systems acquire  steady states with properties strikingly different from equilibrium ones. Three distinct non-stationary  states have been predicted -- a state where amplitude of the BCS order parameter $\Delta(t)$  oscillates periodically \cite{Levitov2004,Levitov2006}, a state with a constant but reduced gap, and a gapless superfluid state [\citealp{Classify}--\citealp{Emil2006}].  Realization of a particular steady state  is determined by the magnitude of change of the pairing strength. Most previous studies concentrated on the time evolution of the order parameter, while direct experimental manifestations of the non-adiabatic dynamics have not been sufficiently explored. The purpose of the present paper is to address this issue.

Amongst existing experimental techniques the RF spectroscopy appears to have the greatest potential for distinguishing different dynamical states  from
equilibrium phases. This motivates us to study  spectroscopic signatures of the dynamics of fermionic condensates.
Our main findings are as follows. In contrast to the BCS ground state spectrum which has a single peak at a frequency determined by the equilibrium gap, the RF spectrum of a nonequilibrium state with
constant but finite  $\Delta_s$ displays two distinct peaks. The second peak reflects the fact that this nonequilibrium state is a superposition of  an infinite number of excited stationary states of the condensate. Excited states contain a mixture of ground state pairs and excited pairs --  two-particle excitations  of the condensate that conserve the total number of
particles {\it and} Cooper pairs (see \cite{BCS,Anderson1958p} and the discussion below).

\begin{figure}[h]
\includegraphics[width=2.8in]{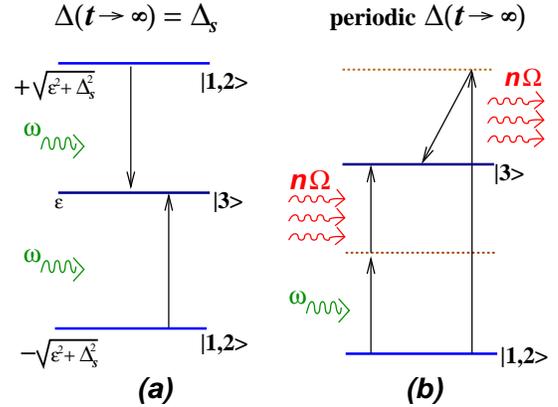}
\caption{(a) Non-stationary state of a fermionic condensate with time-independent order parameter $\Delta_s$ contains a mixture of ground state and excited Cooper pairs of atoms $|1\rangle$ and $|2\rangle$ with energies $\pm\sqrt{\eps^2+\Delta_s^2}$. An RF photon can break either type of pair and transfer one of the atoms to state $|3\rangle$. (b) In the steady state where the order parameter $\Delta(t)$ oscillates with frequency $\Omega$, the photon can  break  a pair and emit/absorb several oscillation quanta.}
\label{Fig1}
\end{figure}

The ``ordinary'' peak present already in the ground state is due to a process whereby  a photon breaks a ground state pair, while in the process responsible for the second peak it breaks an excited pair, (Fig.~\ref{Fig1}(a)). It is interesting to note that in electronic superconductors excited pairs carry no charge or spin and are therefore  difficult to detect.

%%%%%% This is Fig. 1 -> setup to probe non-stationary SC %%%%%%
%%%%%% Done with Fig.1 %%%%%%%%%%%%%%%%%%%%%%%

For a steady state with periodically oscillating order parameter, we show that each of the peaks described above acquires equidistant satellite
peaks, i.e. there are two series of equally spaced peaks in this state. The spacing between peaks in each series is equal to the frequency of oscillation $\Omega$. Satellite peaks appear because a photon can gain optimal energy for breaking a ground/excited pair by emitting or absorbing several ``deltons'' -- oscillation quanta  of energy $\Omega$,   (Fig.~\ref{Fig1}(b)).

In an atomic Fermi gas the pairing occurs between
atoms in two hyperfine states $|1\rangle$ and $|2\rangle$.
The frequency of external RF radiation can be tuned to induce
transitions between one of these states, say $|2\rangle$,
to the third atomic state $|3\rangle$. The  RF spectrum corresponds to the rate of loss of  atoms from $|2\rangle$, i.e.  $I(\omega_{\mbox{\small rf}})=-d{N}_2/dt$, measured as a function of the radiation
frequency $\omega_{\mbox{\small rf}}$. In the normal
state of atoms $|1\rangle$ and $|2\rangle$ the quantity $I(\omega_{\mbox{\small rf}})$ has a sharp peak at $\omega_{\mbox{\small rf}}=\omega_a$, the frequency of atomic transition between  $|2\rangle$ and $|3\rangle$. In the paired ground state the peak shifts
to a larger frequency since now an additional energy is required to break pairs\cite{Chin2004,Torma2000}.

We start with the Hamiltonian $\hat{H}=\hat{H}_{12}^{BCS}+\hat{H}_3+\hat{H}_{23}^{EM}$, where
\begin{equation}
\hat{H}_{12}^{BCS}=\sum\limits_{j,\alpha=1,2}
\varepsilon_j\hat{c}_{j\alpha }^\dagger\hat{c}_{j\alpha }-
\frac{\lambda(t)}{\nu_F}\sum\limits_{ i,j}\hat{c}_{i1 }^\dagger\hat{c}_{i2}^\dagger
\hat{c}_{j2 }\hat{c}_{j1}
\label{Hbcs}
\end{equation}
 is the BCS Hamiltonian describing pairing between  states $|1\rangle$ and $|2\rangle$,  $\hat{c}_{j\alpha}$  ($\alpha=1,2$)  annihilate  atoms in states $|1\rangle$ and $|2\rangle$, $\varepsilon_j$  are  single-particle energy levels relative to the Fermi level of atoms $|1\rangle$ and $|2\rangle$, $\lambda(t)$ and $\nu_F$ are the dimensionless coupling and the density of states at the Fermi level, $\hat{H}_3=\sum_j\varepsilon_j\hat{d}_j^\dagger\hat{d}_j$, where  $\hat{d}_j$ annihilate  atoms in states $|3\rangle$, represents non-interacting atoms in states $|3\rangle$, and
\begin{equation}
\begin{split}
\hat{H}_{23}^{EM}=&\frac{\omega}{2}\sum\limits_{j}(\hat{c}_{j2}^\dagger\hat{c}_{j2}-
\hat{d}_{j}^\dagger\hat{d}_{j})+\hat H_T\\&\hat H_T=\sum\limits_{jl}
(T_{jl}\hat{c}_{j2 }^\dagger\hat{d}_{l}+\text{h.c.}).
\end{split}
\label{H}
\end{equation}
accounts for the interaction of atoms $|2\rangle$ and $|3\rangle$ with the RF
radiation field\cite{Torma2000} in the rotating wave
approximation\cite{RWA}. Here
$\omega=\omega_{\mbox{\small rf}}-\omega_a$ is the detuning frequency.   Since the size of the trap is much smaller than the photon
wavelength, one can take the tunnelling matrix to be diagonal, $T_{jl}=T\delta_{jl}$.

We assume that the pairing strength has been switched from $\lambda_i$ to $\lambda_f$ and the RF radiation is turned on after the condensate has reached
one of the steady states described above. 
The magnitude of the change in  pairing strength is denoted by the parameter $\beta$:
$$
\beta=\lambda_i^{-1}-\lambda_f^{-1}.
$$
Our task is to evaluate the  current  $\langle\hat{I}\rangle=-d{\langle\hat{N_2}\rangle}/dt$. The wave function of the condensate in the steady state without the RF field is of the BCS form $|\Psi(t)\rangle=\prod_j[v_j(t)+u_j(t)\hat{c}_{j1}^\dagger\hat{c}_{j2}^\dagger]|0\rangle$ \cite{Levitov2004}.
Treating the tunnelling  Hamiltonian, $\hat H_T$ in eq.~(\ref{H}), as a perturbation,
we obtain the current out of state $|2\rangle$ to the lowest nonvanishing order in $T_{jl}$
\begin{equation}
%\begin{split}
I= |T|^2\!\!\!\int\limits_{-\infty}^{\infty}d\tilde\omega\!\!\! \sum\limits_{\varepsilon_j\geq\delta\mu}
\!\!\mbox{Re}[ u_j(\varepsilon_j-\omega-\tilde\omega)\overline{u}_j(\omega-\varepsilon_j)e^{i\tilde\omega t}],
%\end{split}
\label{It}
\end{equation}
where $u_j(\omega)$ are  Fourier components of $u_j(t)$ and $\delta\mu=\mu_3-\mu_2$ 
is a difference between the corresponding chemical potentials for atoms in states 
$|3\rangle$ and $|1\rangle$, $|2\rangle$. In recent experiments all states $|3\rangle$ were initially unpopulated \cite{Chin2004,Schunck2007},  which suggests that we set $\delta\mu\simeq{-E_F}$.
However, our model based on truncated BCS Hamiltonian (\ref{Hbcs}) becomes 
invalid for that case since the so-called "off-diagonal" interaction terms
between atoms in states $|1\rangle$ and $|2\rangle$ can not be discarded\cite{Aleiner}. To circumvent
this problem in what follows we assume $|\delta\mu|\ll E_F$. 

%%%%%% This is Fig. 2 -> absorption spectra for const gap %%%%%%
\begin{figure}[t]
\vspace{-2mm}
$\begin{array}{c}
\includegraphics[width=2.8in]{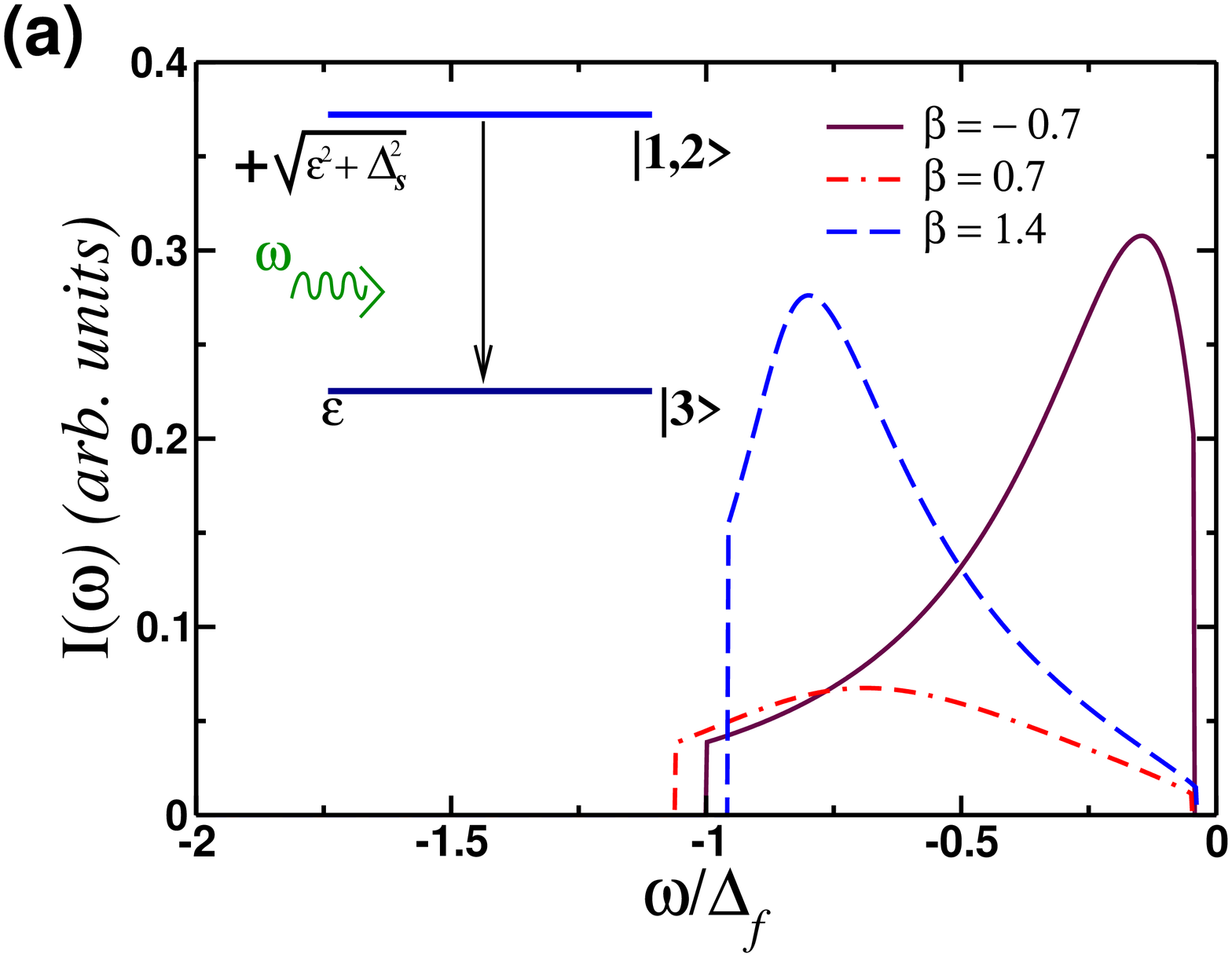}\\
\includegraphics[width=2.8in]{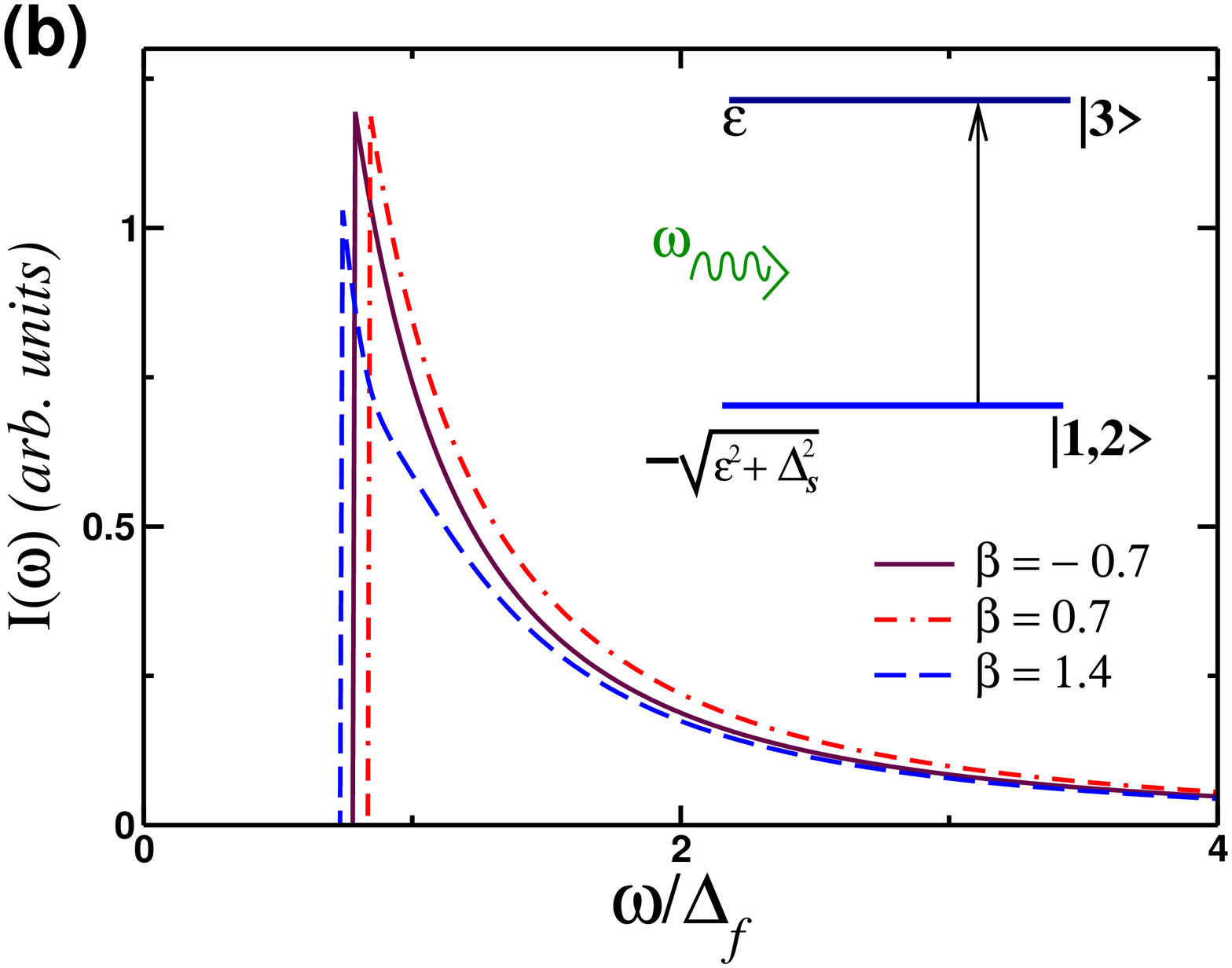}
\end{array}$
\caption{RF spectra (\ref{IBw})  for a
non-stationary  state with a constant order parameter $\Delta_s\ne0$ produced by an abrupt change in the pairing strength, $\lam_i\to\lam_f$ for $\delta\mu=-0.75\Delta_f$; (a) The spectral weight for 
$\omega<0$ where the peak is due to processes where the photon breaks an excited pair. This peak is absent in the ground state, $\beta=0$; 
(b) The peak at $\omega>0$ is due to processes where an RF photon breaks a ground state Cooper pair. A similar peak is present in the paired ground state. $\Delta_f$ is the equilibrium gap 
for the final coupling $\lam_f$}
\label{Fig2}
\end{figure}
%%%%%% Done with Fig.2 %%%%%%%%%%%%%%%%%%%%%%%

\begin{figure}[t]
\includegraphics[width=2.8in]{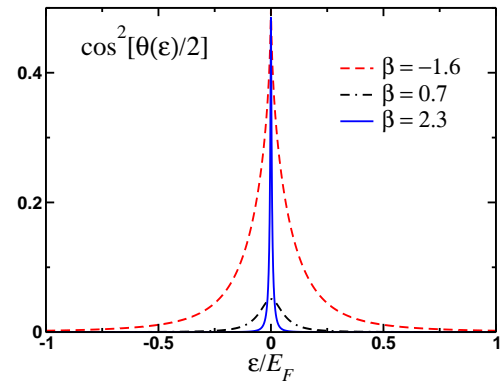}
\caption{The probability $n(\eps)=\cos^2[\theta(\epsilon)/2]$ of having an excited pair (see the text below eq.~\re{bog23}) at energy $\eps$ in all steady states produced by a switch of the BCS coupling constant $\lam_i\to\lam_f$ \cite{Emil2006}. Plots for three  values of $\beta=1/\lambda_i-1/\lambda_f$ are shown.  The presence of excited pairs leads to additional peaks in RF spectra
shown in Figs.~\ref{Fig2},\ref{Fig4}. In the ground state
 $n(\eps)\equiv0$.  }
\label{Fig3}
\end{figure}

Consider first the steady state with a constant order parameter $\Delta_s$ that is realized for
$-\pi/2\le\beta\le \pi/2$. The steady state 
wave function has been determined exactly in Ref.\cite{Emil2006}
\beg
%\begin{split}
\left[\begin{array}{ll}
u_j\\
v_j\\
\end{array}\right]
=\sin\frac{\theta_j}{2}
\left[\begin{array}{ll}
u_j^0\\
v_j^0\\
\end{array}\right]
e^{i\xi_j t}
+
\cos\frac{\theta_j}{2}
\left[\begin{array}{rr}
\bar v_j^0\\
-\bar u_j^0\\
\end{array}\right]
e^{-i\xi_j t+i\phi_j},
%\end{split}
\label{bog23}
\en
where $\xi_j=(\varepsilon_j^2+\Delta_s^2)^{1/2}$, $\phi_j$ is the time-independent relative phase, and $u_j^0=(\xi_j-\varepsilon_j)/2\xi_j$ and $v_j^0=(\xi_j+\varepsilon_j)/2\xi_j$
are the Bogoliubov amplitudes in the BCS ground state with gap $\Delta_s$.
The distribution function 
$\cos^2[\theta(\epsilon_j/2)]$, (Fig.~\ref{Fig3}), and $\Delta_s$ are known exactly in terms of the initial
and final equilibrium BCS gaps $\Delta_i$ and $\Delta_f$ \cite{Emil2006}.
The first term in eq.~\re{bog23} is the wave function of a ground state pair of energy $-\xi_j$. 
The second term is the wave function of an excited pair with energy $\xi_j$ \cite{BCS}. 
Excited pairs are  excitations {\it of} the condensate and should be contrasted 
to the single-particle excitations, which are created {\it outside} of the condensate.
%Excited pairs are elementary excitations {\it of} the condensate as opposed 
%to single-particle excitations {\it %out of} the condensate. 
When the BCS wave function is projected onto the subspace of fixed particle
number \cite{Anderson1958}, excited pairs conserve the total number of paired atoms, while quasiparticle excitations break Cooper pairs. In the Anderson pseudospin representation\cite{Anderson1958}, excited and ground state pairs correspond to a pseudospin respectively aligned parallel or antiparallel to its effective magnetic field. In this case $\theta_j$ is the angle between the pseudospin and the field.

Using eqs.~(\ref{It},\ref{bog23}), we derive the rate of loss in state $|2\rangle$
\begin{equation}
\begin{split}
\frac{I(\omega)}{2\pi|T|^2 }=&\frac{\Delta_s^2}{\omega^2}\left[\sin^2\frac{\theta(\overline{\omega})}{2}
\vartheta\left(\omega-\omega_T^{+}\right)\right.\\& \left.+
\cos^2\frac{\theta(\overline{\omega})}{2}\vartheta(\omega+\omega_T^{-})
\right],
\end{split}
\label{IBw}
\end{equation}
where $\omega_{T}^{\pm}=\sqrt{\delta\mu^2+\Delta_s^2}\pm\delta\mu$
and $\overline{\omega}=(\omega^2-\Delta_s^2)/2\omega$. 
The first term represents the contribution of ground state pairs,
(Fig. \ref{Fig2}(b)), corresponding to a process
where a photon breaks a ground state pair and creates an unpaired atom in state $|3\rangle$, (Fig.~\ref{Fig1}(a)). Energy balance yields $\omega =\eps_j+\xi_j$.  The first term is nonzero when $\omega$ exceeds the threshold energy $\omega_T^{+}$. In the ground state
$\theta(\omega)\equiv \pi$ and only this term remains. The second term derives from excited pairs and corresponds to the process where a photon breaks an excited pair, (Fig.~\ref{Fig1}(a)).  The energy balance now implies $\omega=\eps_j-\xi_j$, which is negative for all $j$. We see that an additional peak appears at $\omega\geq-\omega_T^{-}$, (Fig. \ref{Fig2}(a)). The maximum in absorption is reached at $\omega\approx-\Delta_s$. Its height is suppressed, since $\theta(\omega)$ can only deviate significantly from $\pi$ in an narrow window of width $\Delta_s$ around the Fermi energy, where there is a significant density of excited pairs. Finally, we note that when $\Delta_s\to0$  the two peaks merge at zero frequency, i.e. the RF spectrum of the gapless steady state is reversed to that of a normal state.

Now let us turn to the regime of periodically oscillating order parameter, which occurs when
 $\beta>\pi/2$ \cite{Levitov2004,Levitov2006}. In this state $\Delta(t)$ is given by the Jacobi elliptic function dn with an amplitude comparable to $\Delta_f$ and a period of order $2\pi/\Delta_f$. We are to analytically determine the Bogoliubov amplitudes using the exact solution for the BCS dynamics \cite{Classify,Emil1}, yielding:
\begin{equation}
\begin{split}
\left[
\begin{array}{ll}
u_j \\
v_j\\
\end{array}
\right]
=&\sum_{n=-\infty}^\infty\left(\sin \frac{\theta_j}{2}
\left[
\begin{array}{ll}
a_{jn} \\
b_{jn}\\
\end{array}
\right]
e^{i(\nu_j-n\Omega)t}\right.\\&\left.+
\cos \frac{\theta_j}{2}
\left[
\begin{array}{rr}
\bar b_{jn} \\
-\bar a_{jn}
\end{array}
\right]
e^{-i(\nu_j-n\Omega) t}\right),
\end{split}
\label{unvn}
\end{equation}
where $\Omega$ is the frequency of oscillations of $\Delta(t)$,  $\nu_j=\nu(\eps_j)$ is a function
of single-particle energy, and $\theta_j$ has been discussed below eq.~\re{bog23}. For brevity,  the analytic expressions for $\Omega$, $a_{jn}$,  $b_{jn}$, and $\nu_j$ are omitted. We note however that $\nu_j$ plays analogous role of excitation energy $\xi_j$ for the periodic regime. For example, $\nu_j\to\xi_j$ as
we approach the regime of constant steady state gap, $\beta\to\pi/2$. One can also show
 that $\nu(\eps)$ is a monotonic function of $|\eps|$, $\nu(\eps)\ge|\eps|$, $\nu(0)=\Omega/2$, and $\nu(\eps)\to |\eps|$ for $\lam_i\to0$ and for large $|\eps|$.

Comparison of the steady state wave functions \re{unvn} and \re{bog23} suggests the two terms in eq.~\re{unvn} may be interpreted as two orthogonally paired states for each level $j$. These are the analog of ground state and excited pairs and have energies $\pm \nu_j$.
In addition, these states contain $n$ quanta of the oscillating pairing field $\Delta(t)$ each 
carrying energy $\Omega$. We will refer to these quanta as ``deltons''. 
These are quanta of the amplitude mode of the pairing field and can be interpreted as Higgs bosons \cite{Varma2002,Barankov2007}.

%%%%%% This is Fig. 3 -> cos^2 %%%%%%
%%%%%% Done with Fig.3 %%%%%%%%%%%%%%%%%%%%%%%

\begin{figure}[t]
\includegraphics[width=2.8in]{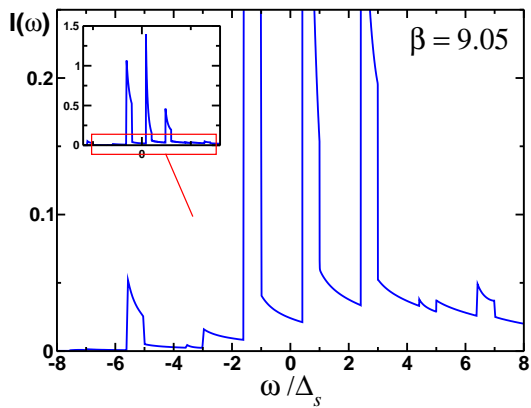}
\caption{RF-absorption spectra eq.~\re{Iwn}  for the state with periodic in time order parameter $\Delta(t)$ produced by a sudden switch of the pairing strength $\lam_i\to\lam_f$ for 
$\delta\mu\simeq-0.75\Delta_s$. The detuning frequency $\omega$ is in units of time-averaged order parameter $\Delta_s$. Note two sequences of peaks at even and odd multiples of $\Delta_s$ and also jumps at $\omega=(2n+1)\Delta_s$ that sometimes are on top of the odd peaks. The frequency of oscillations of $\Delta(t)$ is $\Omega=2\Delta_s$. Multiple peaks are due to processes where an RF photon breaks an excited/ground state Cooper pair and emits or absorbs several oscillation quanta (``deltons''). }
\label{Fig4}
\end{figure}

Equations ~(\ref{It},\ref{unvn}) determine the RF spectrum in the periodic regime
\begin{equation}
\begin{split}
\frac{I(\omega)}{2\pi|T|^2}=&\!\!\!\sum\limits_{n,\epsilon_j\ge\delta\mu}\!\!\!\left\{
\sin^2\frac{\theta_j}{2}|a_{jn}|^2
\delta(\omega-\nu_j-\epsilon_j-n\Omega)\right.\\
&\left.+\cos^2\frac{\theta_j}{2}|b_{jn}|^2
\delta(\omega+\nu_j-\epsilon_j-n\Omega)\right\}.
\end{split}
\label{Iwn}
\end{equation}
Here we dropped oscillatory terms  assuming they  average to zero on the time scale of the measurement. Expression \re{Iwn} describes two series of equidistant peaks, (Fig.~\ref{Fig4}), corresponding to the processes where an RF photon breaks one of the two paired states on level $j$ and emits or absorbs $n$ deltons, (Fig.~\ref{Fig1}(b)). The energy balance reads $\omega=\eps_j\pm\nu_j+n\Omega$.  
The first series of peaks is described by the first term in eq.~\re{Iwn} and is analogous to the ground state pair peak in eq.~\re{IBw}. In this case, the $n=0$ peak is located at the minimum detuning frequency $\omega_T^{+}=\nu(\delta\mu)+\delta\mu$, cf. eq.~\re{IBw}. Thus,  peaks in the first sequence are at $\omega=\omega_T^{+}+n\Omega$. When $\Delta(t)$ is the Jacobi elliptic function dn, 
$\Omega=2\Delta_s$\cite{Ryzhik1965}, 
where $\Delta_s$ is the time average of $\Delta(t)$ over the period.
%for which $\Omega=2\Delta_s$\cite{Ryzhik1965}
%where $\Delta_s$ is the time average of $\Delta(t)$ over the period and 
%we used the fact that $\Delta(t)$ is the Jacobi elliptic function dn, 
%for which $\Omega=2\Delta_s$\cite{Ryzhik1965}. 
We note that
the Fourier components of Bogoliubov amplitudes, $a_{jn}$ and  $b_{jn}$ in eq.~\re{Iwn}, are discontinuous at the Fermi level $\eps_j=0$ similar to  $T=0$ Fermi distribution. This is a consequence of the fact that initial states for the periodic regime are close to the normal state\cite{Levitov2004,Classify,Levitov2006}. The discontinuities lead to jumps in the RF spectra at $\omega=(2n+1)\Delta_s$, (Fig.~\ref{Fig4}).

The second series of peaks is the analog of the excited pair peak in eq.~\re{IBw}. These peaks are at $\omega\approx-\Delta_s+n\Omega=(2n-1)\Delta_s$, (Fig.~\ref{Fig4}). Their heights are suppressed for the same reason as in the excited pair peak  in eq.~\re{IBw}. Their width is determined by the width of the excited pair  distribution function and becomes extremely narrow for large $\beta$, (Fig.~\ref{Fig3}). 
In this limit, they can be superimposed by jumps in the first sequence of peaks, (Fig.~\ref{Fig4}).

%%%%%% This is Fig. 4 -> absorption spectra in Regime A %%%%%%
%%%%%% Done with Fig.4 %%%%%%%%%%%%%%%%%%%%%%%

The sharp features of the RF spectra detailed above will be broadened 
by variety of effects in practice such as changes in particle number between 
experiments. More significant deviations will occur as one gets closer to the 
Feshbach resonance as our treatment is based on BCS theory. 
Finally, RF probing should be performed on a timescale shorter than the 
quasiparticle relaxation time $\tau_\varepsilon\simeq E_F/\Delta_f^2$ \cite{Levitov2006},
which limits the lifetime of the steady states considered here. 
At times larger than $\tau_{\varepsilon}$ an isolated system is
expected to re-thermalize to a state with a nonzero effective temperature which can be determined by balancing the total internal energy \cite{Levitov2006,Emil2006}.

In conclusion, we have obtained RF spectra for the nonequilibrium steady states formed in a fermionic condensate due to a rapid switching of the pairing strength. The RF spectrum of the steady state with constant order parameter $\Delta_s\ne0$ has two peaks in contrast to the spectrum of the paired ground state where there is a single peak. The  peak at negative detuning frequencies reflects the presence of excited pairs -- elementary excitations of the condensate and its shape is a direct measure of their distribution function. The other peak is a counterpart of the ground state spectroscopic response.  In the steady state characterized by a periodically oscillating $\Delta(t)$, each of the two peaks splits into a sequence of equidistant peaks with the spacing between peaks given by the
frequency of oscillations $\Omega$.

The work of M.D. and P. C. was supported by
the DOE grant DOE-FE02-00ER45790. E.Y. was  supported by Alfred P. Sloan Research
Fellowship and NSF grant NSF-DMR-0547769.

\end{document}